\documentclass[prl,aps,twocolumn]{revtex4}
\usepackage[pdftex]{graphicx}
\def\rp{\mathbf{r}_{\perp}}
\def\lsim{\:\raisebox{-0.5ex}{$\stackrel{\textstyle<}{\sim}$}\:}

\newcommand{\beq}{\begin{equation}}
\newcommand{\eeq}{\end{equation}}
\newcommand{\bea}{\begin{eqnarray}}
\newcommand{\eea}{\end{eqnarray}}
\begin{document}
\title{Instabilities and waves in thin films of living fluids}
\author{Sumithra Sankararaman}
\email{sumithra@physics.iisc.ernet.in}
\author{Sriram Ramaswamy\footnote{also at JNCASR, Bangalore 560 064 India}}
\email{sriram@physics.iisc.ernet.in}
\affiliation{Centre for Condensed Matter Theory, Department of Physics, Indian Institute of Science, Bangalore 560012, India}
\begin{abstract}
We formulate the thin-film hydrodynamics of a suspension of polar self-driven particles and show that 
it is prone to several instabilities through the interplay of activity, polarity and the existence of a free surface. Our approach extends, to self-propelling systems, the work of Ben Amar and Cummings [Phys Fluids {\bf 13} (2001) 1160] on thin-film nematics. Based on our estimates the instabilities should be seen in bacterial suspensions and the lamellipodium, and are potentially relevant to the morphology of biofilms. We suggest several experimental tests of our theory. 

\end{abstract}

\maketitle

%The equations of 
Active hydrodynamics, the collective behaviour of self-driven, orientable particles in a fluid medium, is a topic of intense current research \cite{tonertu,reviews,sriram1,cristina,curiegrp,others,bacteria,kaiser}.
%, and apply over a wide range of length scales, from fish shoals to
%bacteria to suspensions of actively contracting filaments in the cytoskeleton.
%This approach to the mechanics of living matter has been pursued vigorously
%over the past few years \cite{others}. 
%Bacterial suspensions \cite{bacteria,kaiser} are a natural arena for testing
%active hydrodynamics. 
In this Letter we study the dynamics of a
film of fluid, thin in the $z$ direction and spread on a solid surface in the
$xy$ plane which is also the easy plane for the orientational order parameter
$\mathbf{p}$ of the {\em polar} active particles suspended in the fluid. 
For an active system, this polarity implies a current $v_0 c \mathbf{p}$ with respect to the fluid, where $v_0$ is a characteristic drift velocity and $c$ the concentration of active particles. While our formulation is general, we study mainly the properties of perturbations about an ordered, uniform  reference state $\langle \mathbf{p} \rangle = \hat{\mathbf{x}}$. 

\begin{figure}[htp]
\centering
\includegraphics[scale=0.5]{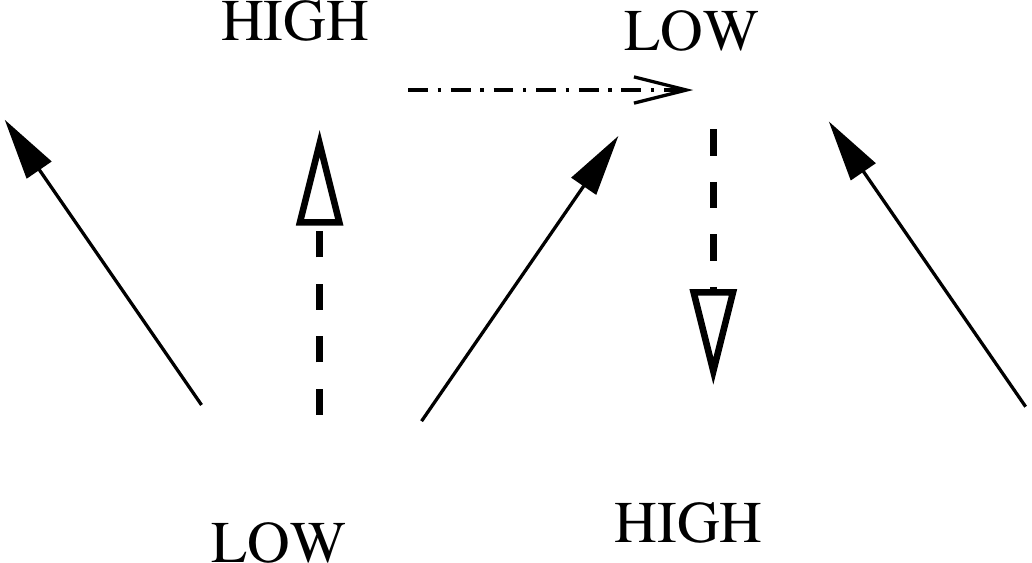}
\caption{Top view of film with contractile polar filaments (solid arrows) with a preference for pointing downhill. Active stresses cause fluid flow (dashed arrows) towards the open end of a splay perturbation, leading to gradients in the height of the fluid film. Modes propagating forward experience a height gradient that further torques (dash-dotted arrow) the filaments in the direction in which they were already perturbed.}
\label{instabfig}
\end{figure}
Our main results are as follows: (i) Active, ordered thin films, although
dominated by viscosity, not inertia, show a {\em wavelike} response to external
disturbances, as a result of the coupled dynamics of free-surface undulations
the active stress field, and the concentration. (ii) In large regimes of parameter space this 
coupling produces a novel instability whose growth rate, to leading order in
small wavevector $\mathbf{k} = (k_x,k_y)$, varies as $k_y k_x^{1/2}$ for $k_x \ll
\xi k_y^2$ and as $k_y^2$ for $k_x \gg \xi k_y^2$. The crossover length $\xi =
\Gamma |C \sigma_0| h_0^2/v_0^2 \mu$ depends on the drift velocity $v_0$ of the
active particles, the mean film thickness $h_0$, the viscosity $\mu$, an orientation mobility $\Gamma$, the coupling $C$, present in equilibrium systems as well, of particle orientation to free-surface tilt, and the typical active contractile or tensile
stress $\sigma_0 \equiv W c_0$. $W$ is the strength of the force-dipole associated with each active particle, and $c_0$ the mean concentration of active particles. 
(iii) Low motility, i.e., small $v_0$, promotes the instability and high motility suppresses it, a point we will return to at the end of this paper. The instability manifests itself in moving (convective) and static (absolute) form.
%, arises only for $k_x$ and $k_y$ both nonzero. 
We also offer a simple physical picture (Fig. \ref{instabfig}) of the mechanism underlying this unique instability. (iv) We find several other possible modes of free-surface instability, which we discuss towards the end of the paper. 

Before presenting our model in detail, a few words are in order about the
systems to which it might apply. Bacterial colonies on a surface frequently
take the form of organised, rigid \textit{biofilms} \cite{biofilmwiki}, e.g.,  those of {\it Pseudomonas aeruginosa} \cite{klausen}.
%During
%formation, a biofilm can be thought of as a thin, ordered, fluid layer of
%active particles, and hence a test-bed for the ideas presented in this paper.
%The biofilms of wild-type {\it Pseudomonas aeruginosa} \cite{klausen} are normally flat 
%carpets, while those of mutants with decreased motility are have irregular, localized bumps. 
An understanding of the formation of these structures should begin with the thin-film hydrodynamics of active fluids. 
The lamellipodium of crawling cells is propelled by the ATP-driven force of a continually 
polymerizing and depolymerizing, contractile actin network just below the cell membrane 
\cite{kruseetalphysbiol} and is thus a moving, thin, active fluid film.

We therefore consider a
fluid film containing active particles with concentration field $c$, and
orientation described by a vector field $\mathbf{p} = (\mathbf{p}_{\perp},p_z)$. Since we are interested here in perturbations about a macroscopically ordered state, we set $|\mathbf{p}| = 1$. We construct equations of motion for $c(\rp, t)$, $\mathbf{p}(\rp, t)$, and the height field $h(\rp,t)$, i.e., the film thickness, as functions of in-plane position $\rp$ and time $t$.  
Our treatment generalizes \cite{benamarcummings} to the case of active systems. 

The kinematic boundary condition \cite{stone} $ \dot h = u_z - \mathbf{u_\perp}\cdot \mathbf{ \nabla_\perp} h$ connects $h$ to the hydrodynamic velocity field $\mathbf{u} =
(\mathbf{u}_\perp, u_z)$ evaluated at the free surface. Incompressibility
$\mathbf{\nabla} \cdot \mathbf{u}$ = 0 leads to the simplification
\begin{equation}
\label{hincompeqn}
\partial_t h + \mathbf{\nabla}_\perp \cdot \int_0^h \mathbf{u}_\perp dz= 0.
\end{equation}
We eliminate 
$\mathbf{u}_{\perp}$ in favour of $\mathbf{p}$, $c$ and $h$ through the Stokes equation in the
%To compute ${\bf u _\perp}$, we use the
lubrication approximation \cite{stone} $u_z=0$, $|\nabla_{\perp} \mathbf{u}|
\ll |\partial_z \mathbf{u}|$:
\begin{equation}
\label{fluideom}
\mu \partial_z^2 \mathbf{u}_\perp -
\mathbf{\nabla}_\perp P - \hat{z} \partial_z P
%- \hat{z} \rho g
-\mathbf{\nabla} \cdot \mathbf{\sigma^a} = 0
\end{equation}
where $\mu$ and $P$ are the viscosity and pressure field of the fluid, and
$\sigma^a(\mathbf{r}) = W c(\mathbf{r}) \mathbf{p}(\mathbf{r})
\mathbf{p}(\mathbf{r})$ is the intrinsic stress field \cite{sriram1,sriram2}
%arising from the permanent force dipole of strength $W$ 
of the active particles.
%and $c(\mathbf{r})$ the concentration of active particles. 
We discuss the role of gravity later in the
paper. $W<0$ and $W>0$ correspond
respectively to contractile and tensile activity.

%For simplicity, we assume as in \cite{tonertu} that the local velocity of
%active particles relative to the fluid is proportional to $\mathbf{p}$. 
The active-particle current is along $\mathbf{p}$, and the particles cannot leave the film. Thus $p_z(z=0) = 0$, and  $\mathbf{p} \cdot {\hat N} = 0$ at $z = h$, where $\hat{N} = (-\mathbf{ \nabla}_\perp h ,1)/\sqrt{1+(\mathbf{\nabla}_\perp h)^2}$ is the outward normal to the free surface. So $p_z \simeq \mathbf{p}_\perp \cdot \mathbf{\nabla}_\perp h$ at the free surface, interpolating linearly, through director elasticity \cite{degp}, to $p_z=0$ at the substrate, i.e., $p_z = (z/h)\mathbf{ p}_\perp \cdot \mathbf{\nabla}_\perp h$. Therefore in a $z$-averaged description $p_z =
(1/2)\partial_x h$ and $\partial_z p_z \simeq h^{-1} \partial_x h$.  Consider
small deviations \cite{footnote}
about a
state aligned along $\hat{\mathbf{x}}$:
$\mathbf{p}_\perp = \hat{x} + \theta \hat{y}$, $\theta \ll 1$.  From the foregoing
discussion, the active force density has components $\nabla_i \sigma_{ix}^a =
\sigma_0 (\partial_y \theta + \partial_x c/c_0 +  h^{-1}\partial_x h)$, $\nabla_i
\sigma_{iy}^a = \sigma_0 \partial_x \theta$ and $\nabla_i \sigma_{iz}^a = \sigma_0 
\partial_x^2 h/2$ to linear order.  We now eliminate the pressure $P$ in favour
of $h$, $\mathbf{p}$ and $c$, through the $z$-component of (\ref{fluideom}) and
stress continuity at the free surface $z=h$, leading to $P(x,y,z,t) = P_0 -
\gamma \nabla_\perp^2 h + \sigma_0 (h-z) \partial_x^2 h/2$, where $P_0$ is a
reference pressure and $\gamma$ the surface tension of the fluid \cite{stone}.
Since we are interested in the dynamics at large scales in the $xy$ plane, we treat the active stresses in a $z-$averaged approximation, allowing us to borrow the methods of \cite{stone} for spreading under gravity, whose force density is $z-$independent. 

Using this in (\ref{fluideom}) and integrating twice over $z$ with 
$\partial_z \mathbf{u}_\perp(h) =0$ and $\mathbf{u}_\perp (0) = 0$
we get
\begin{equation}
\label{uofz} \mathbf{u}_\perp(z) = \frac{hz - z^2/2}{\mu} \left(\gamma
\nabla_\perp \nabla_\perp^2 h 
- {1 \over 2} \sigma_0 h \partial_x^2 \nabla_\perp h- \mathbf{f}_\perp\right), 
\end{equation} 
to linear order in perturbations of $h$, where $\mathbf{f}_\perp = \sigma_0
[(\partial_y \theta + \partial_x c/c_0 +h^{-1} \partial_x h)\hat{x} +
\partial_x \theta \hat{y}]$.  
Using (\ref{uofz}) in (\ref{hincompeqn}), linearizing
$h = h_0 + \delta h$, $c = c_0 + \delta c$, the 
in-plane fourier-components $\delta h_\mathbf{k}(t)$, $\delta c_\mathbf{k}(t)$, 
$\theta_{\mathbf{k}}(t)$ obey 
\begin{eqnarray} 
\label{finalhteqn} 
\partial_t \delta h_\mathbf{k} &=& -\frac{\sigma_0  h_0^2}{3\mu} 
[2h_0k_x k_y \theta_\mathbf{k} + h_0 k_x^2{\delta c_\mathbf{k}
\over c_0} \nonumber \\
&+& (1 - \frac{1}{2} h_0^2 \mathbf{k}^2)k_x^2\delta h_\mathbf{k}]
-\frac{\gamma h_0^3}{3 \mu} \mathbf{k}^4 \delta h_\mathbf{k}.
\end{eqnarray} 

The four effects of activity on the right-hand side
of (\ref{finalhteqn}) are, from left to right within the square bracket, (i) curvature-induced flow; (ii) anisotropic osmotic
flow; and, in parentheses, (iii) splay-induced flow from tilting the free
surface and (iv) an active anisotropic contribution to the effective tension.
The final term on the right of (\ref{finalhteqn}) is ordinary surface tension.
For contractile ($\sigma_0 < 0$) stresses, term (iii) destabilises, and the active
tension in (iv) stabilises, the surface; for tensile ($\sigma_0 > 0$) stresses the
opposite happens. The behaviour of the active tension term is consistent with the idea that contractile stresses pull in along the long axis of the particles, giving additional elastic resistance to stretching along that axis.

The dynamics of the polar orientation field $\mathbf{p}$ differs from that of the conventional nematic director. First, symmetry cannot rule out a coupling of the form 
$-C  \int d^2 x
\mathbf{p}_\perp \cdot \mathbf{\nabla}_\perp h$
%\begin{equation}
%\label{spontsplay1}
 %-C  \int d^2 x
%\mathbf{p}_\perp \cdot \mathbf{\nabla}_\perp h
%\end{equation}
which couples $\mathbf{p}_{\perp}$ to tilts of the free surface \cite{shila}. 
For $C>0$ a tilt of the free 
surface tries to make $\mathbf{p}_{\perp}$ point uphill.

We can think of two possible microscopic origins for such a term. Consider an imposed thickness variation tilt $\mathbf{\nabla}_\perp h$ of the free surface. One end of each polar particle is in general expected to be fatter than the other, and the particles will be best accommodated with the fat end oriented towards the direction of increasing $h$. In addition, the spatial variation in the separation between the free surface and the fixed substrate should give rise to an electric field in the plane. Polar particles in general have an electric dipole moment, and will thus be oriented by this field.  Secondly, consider the spontaneous-splay term \cite{kungetal} 
%\beq
%\label{spontsplay2}
$H_{sp} \equiv -\int
d^3x \tilde{C} \nabla \cdot \mathbf{p} $
%\eeq
in the effective
Hamiltonian for $\mathbf{p}$, where $\tilde{C}$ is a phenomenological parameter which can depend on the local
concentration, say $\tilde{C} = \tilde{C}(c_0) + \tilde{C}'(c_0) \delta c + \ldots 
\equiv C + C' \delta c$. 
Through the ``equilibrium'' part of the dynamics 
of our system, the $C$ and $C'$ terms will contribute $-\Gamma
\delta H / \delta \mathbf{p} = \Gamma C  \delta(z) \nabla_{\perp} h -\Gamma C' \nabla \delta c$ to the equation of motion for $\mathbf{p}$,
$\Gamma$ being a kinetic coefficient. 
\begin{figure}[htp]
\centering
\includegraphics[scale=0.5]{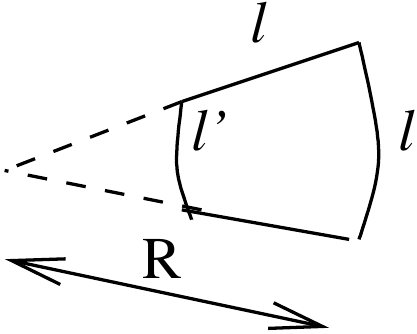}
\caption{An object with head and tail of sizes $\ell$ and $\ell'$ respectively, and length $\ell$, produces a spontaneous curvature $1/R \simeq (\ell - \ell')/\ell^2$.}
\label{spontcurv}
\end{figure}
On dimensional and symmetry grounds we expect $C \sim {\cal F}/R$, where ${\cal F}$ is a typical force scale and $R$, the radius of curvature associated with the particle shape (see 
Fig. \ref{spontcurv})., is the particle thickness divided by the fractional 
length difference between the two edges of the particle.

Combining the above 
%and (\ref{spontsplay1}) 
with the results of \cite{sriram1}, the dynamics of the 
the angle field $\theta$ becomes
\begin{equation}
\label{thetaeqn}
\partial_t \theta 
= -a_1 v_0 \partial_x \theta 
-\zeta \partial_y \delta c
+ \delta(z) \Gamma C \partial_y h 
+ (\lambda A_{yx} - \Omega_{yx})
+ D \nabla^2 \theta.
\end{equation}
The first term on the right-hand side of (\ref{thetaeqn}) is advection of 
$\theta$ 
with a speed proportional to $v_0$ \cite{galinv}. The rest, from left to right, are coupling to concentration gradients, with $\zeta = \Gamma C'$,  coupling to the free surface through %(\ref{spontsplay1}), 
the coefficient $C$, flow-orientation coupling as in nematics \cite{degp}, and gradient elasticity of $\mathbf{p}$. The first three terms are polar, 
and the first is a consequence of activity.  
$A_{ij} = \frac{1}{2} \left( \nabla_i u_j + \nabla_j u_i\right)$ and
$\Omega_{ij} = \frac{1}{2} \left(\nabla_i u_j - \nabla_j u_i \right)$.
For stable flow-alignment
\cite{degp} $|\lambda| > 1$.

Eliminating $\mathbf{u}$ 
in favour of $\theta$ and $h$ through
(\ref{uofz}), and averaging over $z$,
(\ref{thetaeqn}) becomes, to leading orders in $k$ and linear order in the fields,
\begin{eqnarray}
\partial_t \theta &=&  +\frac{i \Gamma C}{h_0} k_y
\delta h_{\mathbf{k}}
- \left(D_+ k_x^2 + D_- k_y^2 + i a_1 v_0 k_x \right) \theta \nonumber \\
&-& (i \zeta k_y - \Phi k_x k_y) \delta c,
\end{eqnarray}
where $ D_{\pm} = D - (\lambda \pm 1) h_0^2\sigma_0 / 4 \mu  $,
and $\Phi =  (\lambda -1) h_0^2 \sigma_0 / 4c_0 \mu  $.

The active contribution to particle currents with respect to the fluid is $v_0 c \mathbf{p}$ where $v_0$ is a drift velocity. Thus in the lab frame of reference the continuity equation for the concentration, apart from diffusion, is
simply $\partial_t \delta c = -\mathbf{\nabla} \cdot [(\mathbf{u} + v_0\mathbf{
p})c]$. Linearizing around $c_0$ and using
(\ref{thetaeqn}) and overall incompressibility,
\begin{equation}
\label{conceom}
\partial_t \delta c = 
-i c_0 v_0 k_y \theta
 - i v_0 k_x \delta c 
+O(k_x^2 \delta c_\mathbf{k}, k_x^2 \delta h_\mathbf{k}).
\end{equation}
The $O(k^2)$ terms in (\ref{conceom}) come from
including fluctuations in $p_x$ as well by explicitly including diffusion of particles. 

The instability we are chiefly concerned with here, arising from the combination of activity and the tilt coupling $C$, is best seen in the extreme case of {\em immotile but active} polar particles, {\it i.e.} $v_0 = 0$ but $\sigma_0 \neq 0$. To leading order in $k$ the concentration  plays no role, and the dynamics is determined by (\ref{finalhteqn}) and (\ref{thetaeqn}), yielding an unstable mode in the system which grows at a rate $k_x^{1/2}k_y$. In detail, for time-dependence $\exp(-i \omega t)$, to leading order in $k$, the mode frequency 
\beq
\label{newinstab}
\omega = \pm {1+i \,\mbox{sgn}(k_x C \sigma_0) \over \sqrt{2}} ({\Gamma h_0^2 \over 3 \mu})^{1/2}|C \sigma_0 k_x|^{1/2} |k_y| 
\eeq
has a real and an imaginary part; the mode displaying an instability travels in the $+x$ direction ($-x$ direction) if $\sigma_0 C >0$ ($\sigma_0 C<0$).

This instability can be understood from a simple physical picture, involving the interplay of active stresses $\sigma_0$ and spontaneous splay $C$. To fix ideas, consider as in Fig. \ref{instabfig} a long-wavelength splay perturbation on a set of {\em contractile} filaments $\sigma_0 < 0$ aligned with polarisation along $+\hat\mathbf{x}$, with a preference for pointing {\em downhill} ($C<0$), in an initially flat film. The active stresses then generate a flow in the film that increases the height of the free surface ahead of the outward-splayed filaments, and decreases the height ahead of the inward-splayed filaments. The mode propagating along $+\hat\mathbf{x}$ will then cause the filaments to sample a height gradient that tilts the filaments in the centre of the field further to the right. The same argument, \textit{mutatis mutandis}, goes through for the tensile and/or $C>0$ cases. 

For motile particles, $v_0 \ne 0$, let us for the moment continue to ignore the concentration field. In the limit $\Gamma C/v_0 \gg 1$, (\ref{finalhteqn}) and (\ref{thetaeqn}) then yield mode frequencies 
\bea
\omega_1 &=& -i{\Gamma C \over v_0}{2\sigma_0 h_0^2 \over 3\mu a_1}[k_y^2 + O({v_0 \over \Gamma C})k_x^2]
\label{modes_vnonzero_noconc1}\\
\omega_2 &=& a_1 v_0 k_x +i{\Gamma C \over v_0}{2\sigma_0 h_0^2 \over 3\mu a_1}[k_y^2 -  O({v_0 \over \Gamma C})D_{\pm} k^2]
\label{modes_vnonzero_noconc2} 
\eea

We now include the concentration via (\ref{conceom}), but continue in the simplifying limit $\Gamma C/v_0 \gg 1$. One mode retains the purely relaxational 
form (\ref{modes_vnonzero_noconc1}), with coefficient modified at $O(\zeta)$. 
The propagating mode in (\ref{modes_vnonzero_noconc2}) becomes a pair with speeds of order $v_0 \pm O(\zeta)$, with relaxation rates $\sim \pm (h_0^2 \sigma_0 C \Gamma / \mu v_0)[1 + O(\zeta)]k^2$.  
The results from (\ref{newinstab}) to this point establish the main claims at the start of this paper. In particular, it is readily seen that (\ref{modes_vnonzero_noconc1}), (\ref{modes_vnonzero_noconc2}) crossover to (\ref{newinstab}) for $k_x \ll \xi k_y^2$ with $\xi = \Gamma |C \sigma_0| h_0^2/v_0^2 \mu$.

Some general features are worth noting: (a) Regardless of the sign of $\sigma_0 C$, there is always an instability; (b) increasing $v_0$ weakens the instability, presumably because a collection of filaments drifting along $x$ samples alternating height gradients along $y$ whose effect cancels; (c) for $\sigma_0 C > 0$ the instability moves with a speed $\sim v_0 \pm O(\zeta)$, combining the effects of the drift speed $v_0$ and the 
``pressure'' due to $\zeta$, while for $\sigma_0 C < 0$ there is an instability in mode $\omega_1$, predominantly the height field $h$, that grows without travelling, the remnant of the instability discussed for $v_0 =0$. 

It is crucial to note here that \textit{activity} and \textit{polarity} conspire to produce the instability, at leading order in gradients, even at zero \textit{motility}. Motility, also a consequence of active polarity, is encoded in the active stress only at next-to-leading order \cite{sriram2}. The stabilising effect of increasing $v_0$ \textit{at constant $\sigma_0$} is thus not paradoxical. 

For $\Gamma C/v_0 \ll 1$ all instabilities involving the interplay of activity with the spontaneous-splay coupling $C$ disappear. The remaining instabilities involve neither polarity nor a drift velocity. The most interesting of these, mentioned briefly below (\ref{finalhteqn}), arises as follows: Consider a flat free surface with {\em contractile} filaments aligned along $\hat\mathbf{x}$, and impose a small tilt $\delta h \propto x$. The filaments at the free surface are then tilted relative to those at the substrate. The resulting splay in the $xz$ plane implies that active stresses will pump fluid towards the open end of this splayed  configuration, thus increasing the tilt. 
In addition, the instabilities of a bulk active ordered suspension as originally discussed in \cite{sriram1,curiegrp} can arise, modified here by confinement to a thin film, and the anisotropic active tension [mentioned after (\ref{finalhteqn})], if large enough, will destabilize {\em tensile} active films at order $k^4$. 

We now compare the relaxation rates in (\ref{newinstab}) with the stabilizing contributions $\rho g h_0^3 k^2/\mu$ due to gravity, $\rho$ being the mass-density of the film and $g$ the acceleration due to gravity, $\gamma h_0^3 k^4/\mu$ due to surface tension, and $K k^2/\mu$ due to orientational relaxation, where $K$ is a Frank constant. 
On dimensional grounds and from standard liquid-crystal physics \cite{degp} we take  $\Gamma \approx 1/\mu$, where $\mu \approx 10^{-3}$ Pa s is the viscosity of the medium. With this estimate we find the instability survives these stabilizing agencies provided 
$kh_0 < \min[C \sigma_0/\rho^2g^2h_0^3, (C \sigma_0h_0/\gamma^2)^{1/5}, 
C \sigma_0 h_0^3/K^2]$. For $\gamma$ and $\rho$ we use values for water. $\sigma_0$ should be of order $f a c_0 = \phi f/a^2$ where 
$f$ is the force exerted by the active particle on the fluid, $a$ the particle size, and we take the particle volume fraction $\phi = c_0 a^3$ of order unity since we want an ordered phase. For a bacterium of size $a \sim$ a few $\mu$m moving at several $\mu$m/s through water $f \sim 1$ pN. We argued
% in Eq. (\ref{Ceqn}) and Fig. \ref{spontcurv} 
that $C \sim {\cal F}/R$ where ${\cal F}$ is a force scale and $R$ the spontaneous curvature associated with a polar object. For thermal hard-rod systems ${\cal F} \sim k_BT/\ell$ where $\ell$ is a rod length. For $\mu$m-sized objects this would give $C \sim 10^{-6}$ dyn/cm. For a self-propelled object, it is possibly more reasonable to use the force-dipole strength $W = a f$ as the natural energy scale, in which case ${\cal F} \sim f$. For an object of radius 2 $\mu$m moving through water at 20 $\mu$m/s this gives ${\cal F}$ in the range of a piconewton and thus $C \sim 10^{-4}$ dyn/cm. For film thickness $h_0 \simeq 20$ $\mu$m, and $C \sim 10^{-6}$ dyn/cm, the instability should be seen if $kh_0 \lsim 10^{-3}$ . However, this is a pessimistic estimate: First, activity rather than thermal energy is quite likely responsible for $C$. Secondly, the tension $\gamma$ for biofilms is that between the film and the ambient aqueous medium and therefore much smaller than the air-water value. We expect therefore that the instability should be seen over a much wider wavenumber range, possibly $k h_0 \lsim 1$.

The experiment of choice to test our ideas would be to compare the dynamics of two initially flat thin films of uniform concentration, one with highly motile bacteria, the other with a low-motility mutant, under conditions of constant bacterial concentration. The mutant population would correspond to a system with $v_0$ small, and should show our instabilities. The observations of \cite{klausen} on {\it Pseudomonas aeruginosa} are of interest in this regard, but are complicated by the fact that the bacteria are dividing. The relation of our mode structure to the excitations seen in studies \cite{kaiser} of {\it Myxococcus xanthus} and to the 
lamellipodium of crawling cells \cite{kruseetalphysbiol} remains to be explored. A complete  treatment must include actin treadmilling, as well as the elasticity and anchoring of the cell membrane. In all these examples, the dynamics of the bulk fluid above the film must also be included. Since inhomogeneities in orientation give rise to mass flux, the instabilities will produce large concentration variations which will be important especially when going beyond the linearised treatment. 

We thank  S. Banerjee for pointing out an error in an earlier version of this paper, J. Prost, K. Vijay Kumar and J. K\"as for valuable inputs, and CEFIPRA project 3504-2 and the DST Math-Bio Centre grant SR/S4/MS:419/07 for support.


\begin{thebibliography}{20}
\bibitem{footnote} 
Even in a linearized description,
$p_x$ is not constant but is slaved to the concentration and height
fields, as seen by generalizing the treatment of \cite{tonertu} to
include the effects of the free surface. This results in subdominant
contributions to the dynamics of the height field (\ref{finalhteqn}), which we ignore here.
\bibitem{galinv} The proportionality constant $a_1$ is not necessarily unity. Since the velocities are defined with respect to the fluid's frame of reference, shifting all velocities by a constant is not an invariance \cite{tonertu}.
\end{thebibliography}
\end{document}